\documentclass[aps,pra,graphics,twocolumn,floatfix,notitlepage,superscriptaddress]{revtex4-2}
\usepackage[english]{babel}
\usepackage{textcomp,xcolor,natbib}
\usepackage{dcolumn}    
\usepackage{graphicx}
\graphicspath{{Figures/}}

\usepackage{amsmath, bbm}
\usepackage{latexsym}
\usepackage{amsfonts}   
\usepackage{amssymb}
\usepackage{array}      
\usepackage{epsfig}
\usepackage{txfonts}
\usepackage{xcolor,braket}
\usepackage[colorlinks=true,linkcolor=blue,urlcolor=blue,citecolor=blue]{hyperref}
\usepackage[normalem]{ulem}
\usepackage{soul}
\usepackage{dsfont}

\usepackage{xfrac}  
\usepackage{relsize}

\usepackage{mathrsfs}

\usepackage{cancel}
\usepackage{enumitem}  

\catcode`\>=\active \def>{
\fontencoding{T1}\selectfont\symbol{62}\fontencoding{\encodingdefault}}

\newcommand{\tmop}[1]{\ensuremath{\operatorname{#1}}}
\newcommand{\tmrsub}[1]{\ensuremath{_{\textrm{#1}}}}

\newcommand{\assign}{:=}
\newcommand{\cdummy}{\cdot}

\begin{document}

\title{Holevo skew divergence for the characterization of information
backflow}

\author{Andrea Smirne}
\email{andrea.smirne@unimi.it}
\affiliation{Dipartimento di Fisica “Aldo Pontremoli”, Università degli Studi di Milano, via Celoria 16, 20133 Milan, Italy}
\affiliation{Istituto Nazionale di Fisica Nucleare, Sezione di Milano, via Celoria 16, 20133 Milan, Italy}
\author{Nina Megier}
\affiliation{Dipartimento di Fisica “Aldo Pontremoli”, Università degli Studi di Milano, via Celoria 16, 20133 Milan, Italy}
\affiliation{Istituto Nazionale di Fisica Nucleare, Sezione di Milano, via Celoria 16, 20133 Milan, Italy}
\affiliation{International Centre for Theory of Quantum Technologies (ICTQT), University of Gdańsk, 80-308 Gdańsk, Poland}
\author{Bassano Vacchini}
\email{bassano.vacchini@mi.infn.it}
\affiliation{Dipartimento di Fisica “Aldo Pontremoli”, Università degli Studi di Milano, via Celoria 16, 20133 Milan, Italy}
\affiliation{Istituto Nazionale di Fisica Nucleare, Sezione di Milano, via Celoria 16, 20133 Milan, Italy}

\begin{abstract}
  The interpretation of non-Markovian effects as due to the information exchange
  between an open quantum system and its environment has been recently formulated in terms of
  properly regularized entropic quantities, 
  as their revivals in time 
  can be upper bounded by means of quantities describing the storage of information outside the open 
  system [Phys. Rev. Lett. 127, 030401 (2021)]. Here,
  we elaborate on the wider mathematical framework of the theory,
  specifying the key properties that allow us to associate distinguishability quantifiers
  with the information flow from and towards the open system.
  We point to the Holevo quantity as a distinguished quantum divergence to which the formalism
  can be applied, and we show how several distinct quantifiers of non-Markovianity 
  can be related to each other within this general framework.
 Finally, we apply our analysis to two relevant physical models in
  which an exact evaluation of all quantities can be performed.
\end{abstract}

\maketitle
\section{Introduction}
{The interaction of an open quantum system with its surrounding environment 
will typically result in a non-Markovian dynamics, in which memory effects occur, for example,
due to strong system-environment coupling or low temperature and, more in general, when 
the evolution of the environment takes place on similar time scales as compared to the relaxation time 
of the open system~\cite{Breuer2002,Rivas2012}.}

The physical picture behind non-Markovian dynamics is that the interaction between the open system and the environment 
establishes significant
correlations among them, as well as changes in the environment, which subsequently affect
the evolution of the open system, thus leading to memory effects \cite{Rivas2014,Breuer2016,Vega2017}.
The first quantitative definition of memory effects in open quantum systems has been given in terms of the trace distance \cite{Breuer2009,Laine2010}.
The latter quantifies the distinguishability of quantum states \cite{Nielsen2000}, 
so that its increase in time can be read as due to some information flowing back to the open system, 
leading to an enhanced capability of distinguishing among pairs of open-system states.
In addition, the triangular inequality and the contractivity under completely positive trace preserving (CPTP) maps of the trace distance allows us to link unambiguously 
memory effects, and thus the non-Markovianity quantified by means of it, to the system-environment correlations and the changes in the environment 
during the dynamics~\cite{Laine2010b,Mazzola2012,Smirne2013,Cialdi2014,Campbell2019}.  
{Indeed, this notion of quantum non-Markovianity is referred to open-system evolutions
where the initial system-environment correlations can be neglected,
as the latter would generally prevent 
from the existence of reduced dynamical maps in the first place \cite{Pechukas1994,Alicki1995,Stelmachovic2001}.}
{Other approaches to non-Markovianity have also been considered, e.g. dealing with multi-time measurements \cite{Pollock2018a,Budini2018b,Budini2019a,Yu2019a,Budini2021b,Budini2022a}.}

The possibility to introduce alternative ways to define memory effects, based on different distinguishability quantifiers,
has been investigated from the very beginning \cite{Laine2010}.
Moving from distances to quantum divergences, relative entropy represents indeed a natural candidate, due to its informational meaning 
associated with the optimal strategy to discriminate over two probability distributions in an asymmetric hypothesis testing when an arbitrarily 
large number of measurements is allowed \cite{Bengtsson2017}, as well as due to its contractivity under CPTP maps.
However, the relative entropy can easily diverge, also for finite-dimensional systems, which would lead to singularities in the corresponding
measure of non-Markovianity \cite{Laine2010}.
Even though some entropic quantifiers have been used to define quantum non-Markovianity \cite{Fanchini2014,Haseli2014,Kolodynski2020}, 
only recently \cite{Megier2021} it has been proven that properly regularized versions of the relative entropy \cite{Audenaert2014}
can be equipped with a full interpretation as quantifiers of information backflow, connecting their revivals 
to the microscopic features of the evolution of the open system and the environment.

Here, we further extend this approach, by showing that it is part of a more general mathematical framework, which encompasses
several significant distinguishability quantifiers, including both distances and divergences that are not necessarily distances \cite{Bengtsson2017}. 
First, we present three key properties that guarantee a fully meaningful use of 
distinguishability quantifiers to characterize the exchange of information between the open system and the environment.
These properties allow us to derive in full generality an upper bound to the revivals of the distinguishability
quantifier at hand, linking any backflow of information toward the open system to some information stored within the system-environment correlations or the environment; importantly, the information content both within and outside the open system is defined by means of the same quantifier.
We then show that a normalized version of the Holevo quantity provides us with a significant instance of the general framework, in this way connecting
the very notion of quantum non-Markovianity as information backflow 
to a quantity of primary importance in quantum information theory \cite{Nielsen2000}. 
Moreover, our approach relates distinct quantifiers of non-Markovianity within a common framework, as we show by taking into account a generalized version of the trace distance, which has been investigated extensively in the context of quantum non-Markovianity \cite{Vacchini2011,Chruscinski2014,Wissmann2015,Breuer2016}, and a symmetrized version of the regularized relative entropy considered in \cite{Megier2021}.
Lastly, we evaluate explicitly the behavior in time of the different 
quantifiers of information in simple, but physically relevant examples, illustrating their similarities and
differences.

The rest of the paper is organized as follows. In Sec.\ref{sec:cfn}, we present the general framework 
in which the revivals of any distinguishability quantifier satisfying three definite properties 
are associated with a backflow of information to the open system, by means of a general upper bound to the distinguishability variations
in terms of the information within the system-environment correlations and environmental changes. In Sec.\ref{sec:holevo-skew-diverg}, we show that a normalized
version of the Holevo quantity falls within this framework, and we further provide a tighter upper bound to its variations,
which still keeps the same physical interpretation.
The Helstrom norm of the weighted difference of two quantum states, which includes the trace distance as a special case, and a regularized and symmetrized
version of the relative entropy are considered in Sec.\ref{sec:dad}, where it is also shown the connection to the Jensen-Shannon divergence
for a proper choice of the defining parameters.
Sec.\ref{sec:ex} presents the application of the general analysis to the spin-star system and 
the Jaynes-Cummings model,
while the conclusions of our work are given in Sec.\ref{sec:con}.

\section{Criteria for non-Markovianity quantifiers}\label{sec:cfn}

We start by introducing the general framework we use to define the non-Markovianity of the
dynamics of open quantum systems in terms of information backflow. The main
aim of this construction is to clarify what are the properties a quantifier of
information needs, to capture the physical meaning of information exchange
between an open system and its environment. Crucially, this can be done only
taking into account, besides the information content within the open system itself,
the global system-environment degrees of freedom where the
information can be stored and accessed subsequently in the course of the
open-system evolution.
{We proceed in two steps: in Sec.\ref{sec:dp} we introduce the properties
fixing the class of quantifiers of information we refer to, taking into account their action on
the pairs of open-system states and on their behavior under CPTP maps;
the connection with a CPTP open-system dynamics, under the assumption of an initial product state, 
is then presented in Sec.\ref{sec:ieb}, where the defining properties are linked
to a precise characterization of the information exchanges between the open system and the environment.}

\subsection{Defining properties}\label{sec:dp}

The basic idea is that the changes in the information content of a physical
system can be quantified by looking at how the distinguishability among the
states of the system varies in time {\cite{Breuer2009,Laine2010,Breuer2016}}.
A decrease of the distinguishability indicates a leak of information from the
system at hand to some other degrees of freedom. Conversely, an increase of
the distinguishability means that some information has been recovered by the
system, leading to those memory effects that are at the core of the notion of
non-Markovian quantum dynamics.

The picture now recalled can be formulated in a general and consistent way by
quantifying the distinguishability of any couple of states $\rho$ and $\sigma$
via a quantity $\mathfrak{S} (\rho, \sigma)$ that satisfies the following
properties.
\begin{itemize}
  \item[I.] Boundedness, normalization and indistinguishability of identical
  states:
  \begin{equation}
    \label{eq:bo} 0 \leqslant \mathfrak{S} (\rho, \sigma) \leqslant 1 \quad
    \forall \rho, \sigma,
  \end{equation}
  with
  \begin{equation}
    \label{eq:norm} \mathfrak{S} (\rho, \sigma) = 1 \quad \Leftrightarrow
    \quad {\rho \perp_{\mbox{\footnotesize{supp}}} \sigma},
  \end{equation}
  where {$\rho \perp_{\mbox{\footnotesize{supp}}}\sigma$} means that the two states have orthogonal support
  (e.g., they are orthogonal if they are pure), and
  \begin{equation}
    \label{eq:indid} \mathfrak{S} (\rho, \sigma) = 0 \quad \Leftrightarrow
    \quad \rho = \sigma .
  \end{equation}
  The information stored within a system or exchanged between different
  degrees of freedom is finite and the corresponding quantifier is normalized
  to one. Such a normalization guarantees a fair comparison among different
  quantifiers, 
  as we will see for the relevant example of the Holevo quantity in the next section. 
  In addition, the requirement that two identical states cannot be distinguished,
  while all others can at least to a certain extent, immediately translates
  into Eq.~\eqref{eq:indid}.
  
  \item[II.] Contractivity under CPTP
  maps:
  \begin{equation}
    \mathfrak{S} (\Lambda [\rho], \Lambda [\sigma]) \leqslant \mathfrak{S}
    (\rho, \sigma) \quad \forall \rho, \sigma, \hspace{0.17em} \hspace{0.17em}
    \forall \text{CPTP } \Lambda . \label{eq:cp}
  \end{equation}
  The distinguishability between the states of a quantum system cannot be
  increased by acting locally on the system; as we will see,
  distinguishability can be instead increased if the system is correlated with
  other degrees of freedom. Indeed, this property is strictly related to the
  quantum data-processing inequalities, stating that the information content
  of a quantum system cannot be enhanced via local data processing on that
  system {\cite{Nielsen2000}}.
  
  \item[III.] Triangle-like inequalities:
  \begin{eqnarray}
    \mathfrak{S} (\rho, \sigma) -\mathfrak{S} (\rho, \tau) & \leqslant & \phi
    (\mathfrak{S}(\sigma, \tau)) \quad \forall \rho, \sigma, \tau 
    \label{eq:tlike1}\\
    \mathfrak{S} (\sigma, \rho) -\mathfrak{S} (\tau, \rho) & \leqslant & \phi
    (\mathfrak{S}(\sigma, \tau)) \quad \forall \rho, \sigma, \tau, 
    \label{eq:tlike2}
  \end{eqnarray}
  with $\phi$ a concave function that is strictly positive for a positive
  argument, while $\phi (0) = 0$.
  
  This property generalizes the triangle inequality and it allows us to take
  into account quantifiers of information that are not necessarily distances.
  As we will show in Sec.\ref{sec:ieb}, the triangle-like inequalities are the
  key identities that relate the changes in the information about
  a system to the information content within other degrees of freedom.
\end{itemize}
Broadly speaking, we insist on two classes of objects that satisfy the
properties I.-III.: distances and entropic quantifiers; to include both of
them and to stress that we are referring to quantifiers of state
distinguishability that are not necessarily distances, releasing therefore
symmetry and triangle inequality, we call any quantity satisfying I.-III. a
quantum divergence {\cite{Bengtsson2017}}. We stress that both properties I. and III. set nontrivial constraints
for the case of entropic quantifiers, as it is immediately
clear considering the unboundedness of the standard quantum relative entropy.
On the contrary, while property III. is satisfied by any distance (indeed, in the
form of a proper triangle inequality, with $\phi (x) = x$), note that the same is not true for properties I. and II., as can be seen considering for example the Hilbert-Schmidt distance.

\subsection{Information exchange between an open system and its
environment}\label{sec:ieb}

We now show how any quantum divergence with the abovementioned properties
leads to a consistent characterization of the information flow from and toward
an open quantum system, i.e., a quantum system that is interacting with an
environment.

We assume that the open system $S$ and the environment $E$ are uncorrelated at
the initial time $t_0 = 0$, i.e., $\rho_{SE} (0) = \rho_S (0) \otimes \rho_E
(0)$, with a fixed environmental state $\rho_E (0)$. The evolution of the open
system is thus characterized by a family of CPTP maps $\Lambda (t)$, according
to~{\cite{Breuer2002}}
\begin{equation}
  \rho_S (t) = \Lambda (t) [\rho_S (0)] = \text{tr}_E \{ U (t) (\rho_S (0)
  \otimes \rho_E (0)) U^{\dagger} (t) \},
\end{equation}
where tr\tmrsub{$E$} is the partial trace over the environmental degrees of
freedom and $U (t)$ fixes the unitary global system-environment dynamics. As
said, we want to follow the evolution in time of the distinguishability for
the different degrees of freedom involved, both within and outside the open
system. To do so, we consider two different initial conditions, $\rho_{SE} (0)
= \rho_S (0) \otimes \rho_E (0)$ and $\sigma_{SE} (0) = \sigma_S (0) \otimes
\sigma_E (0)$, with $\rho_E (0) = \sigma_E (0)$, so that the reduced dynamics
is given in both cases by the same family of CPTP maps, $\rho_S (t) = \Lambda
(t) [\rho_S (0)]$ and $\sigma_S (t) = \Lambda (t) [\sigma_S (0)]$. Taking two
instants of time $s$ and $t \geq s$ and using a generic quantum divergence
$\mathfrak{S}$ to quantify distinguishability, the difference
\begin{equation}
  \label{eq:delta} \Delta_S \mathfrak{S} (t, s) \assign \mathfrak{S} (\rho_S
  (t), \sigma_S (t)) -\mathfrak{S} (\rho_S (s), \sigma_S (s))
\end{equation}
tells us the variation in the information content within the open system from
time $s$ to time $t$. Furthermore, $\mathfrak{S}$ can be used to quantify the
information within the environment, looking at $\mathfrak{S} (\rho_E (t),
\sigma_E (t))$ (where $\rho_E (t) = \text{tr}_S \{ \rho_{SE} (t) \}$ is the
environmental state at time $t$), or the information that is shared by the
system and the environment, contained in their correlations and expressed by
$\mathfrak{S} (\rho_{SE} (t), \rho_S (t) \otimes \rho_E (t))$ and
$\mathfrak{S} (\sigma_{SE} (t), \sigma_S (t) \otimes \sigma_E (t))$. The
defining properties II. and III. of quantum divergences imply that the
information variation can always be bounded by
\begin{eqnarray}
  \Delta_S \mathfrak{S} (t, s) & \leqslant & \phi \circ \phi (\mathfrak{S}
  (\rho_E (s), \sigma_E (s)))  \label{eq:main}\\
  &  & + \phi (\mathfrak{S} (\rho_{SE} (s), \rho_S (s) \otimes \rho_E (s)))
  \nonumber\\
  &  & + \phi (\mathfrak{S} (\sigma_{SE} (s), \sigma_S (s) \otimes \sigma_E
  (s))), \nonumber
\end{eqnarray}
where $\circ$ denotes the composition of functions. This relation provides us
with a complete physical interpretation of the changes in the information flow
from and toward the open system, along with their microscopic origin. Any
backflow of information to the open system from time $s$ to time $t$, leading
to the revival $\Delta_S \mathfrak{S} (t, s) > 0$, is due to some information
contained at time $s$ within the environmental degrees of freedom or the
system-environment correlations. In fact, since $\phi (0) = 0$ and due to the
indistinguishability of identical states in (\ref{eq:indid}), the right hand
side (r.h.s.) of Eq.~(\ref{eq:main}) can be different from zero only if at least one of
the following occurs: (i) $\rho_E (s) \neq \sigma_E (s)$, (ii) $\rho_{SE} (s)
\neq \rho_S (s) \otimes \rho_E (s)$, (iii) $\sigma_{SE} (s) \neq \sigma_S (s)
\otimes \sigma_E (s)$. The seemingly trivial fact that a proper
information-flow quantifier $\mathfrak{S} (\rho, \sigma)$ takes the minimum
value equal to zero if and only if $\rho = \sigma$ thus plays quite an
important role in our framework. 
In fact, this condition allows us to conclude that a revival
$\Delta_S \mathfrak{S} (t, s) > 0$ is necessarily due to the presence at time
$s$ of system-environment correlations or changes in the environmental state.
Indeed, this generalizes the corresponding results for the trace distance
{\cite{Laine2010b,Mazzola2012,Smirne2013,Cialdi2014}}, recently extended to a
proper entropic quantifier in {\cite{Megier2021}} (see also
Sec.\ref{sec:dad}). By summing the revivals $\Delta \mathfrak{S}_S (t, s)$
along the whole time evolution (and possibly maximizing over the couple of
initial system states), we can define a quantifier of the non-Markovianity of
quantum dynamics for any quantum divergence $\mathfrak{S}$ exactly in the same
spirit as the one based on trace distance {\cite{Breuer2009,Laine2010}}.
By virtue of Eq.(\ref{eq:main}), memory effects are thus traced back unambiguously to a two-fold exchange of
information, from the open system to the environment and their correlations: a
finite amount of information is stored in external physical degrees of freedom and
later retrieved.

To prove Eq.~(\ref{eq:main}), we first note that the contractivity of
$\mathfrak{S}$ under CPTP maps implies its invariance under unitary maps,
\begin{equation}
  \label{eq:uinv} \mathfrak{S} (U \rho U^{\dagger}, U \sigma U^{\dagger})
  =\mathfrak{S} (\rho, \sigma) \quad \forall \rho, \sigma, \hspace{0.17em}
  \hspace{0.17em} \forall \text{unitary } U,
\end{equation}
as well as under the tensor product with a fixed state,
\begin{equation}
  \label{eq:trinv} \mathfrak{S} (\rho, \sigma) =\mathfrak{S} (\rho \otimes
  \tau, \sigma \otimes \tau) \quad \forall \rho, \sigma, \tau ;
\end{equation}
the former invariance holds since both $U \cdot U^{\dagger}$ and its inverse
$U^{\dagger} \cdot U$ are CPTP maps, while the latter since both the partial
trace and the tensor product with a fixed state are CPTP
maps~{\cite{Megier2021}}. We thus have
\begin{eqnarray}
  \Delta_S \mathfrak{S} (t, s) & \leqslant & \mathfrak{S} (\rho_{SE} (t),
  \sigma_{SE} (t)) -\mathfrak{S} (\rho_S (s), \sigma_S (s)) \\
  & = & \mathfrak{S} (\rho_{SE} (s), \sigma_{SE} (s)) -\mathfrak{S} (\rho_S
  (s), \sigma_S (s)), \nonumber
\end{eqnarray}
where in the first line we used Eq.~(\ref{eq:cp}) (with respect to the CPTP map
tr\tmrsub{$E$}), and in the second line Eq.~(\ref{eq:uinv}) (with respect to
the unitary map $U (s) U^{\dagger} (t)$). Now we sum and subtract
$\mathfrak{S} (\rho_S (s) \otimes \rho_E (s), \sigma_{SE} (s))$, and replace
$\mathfrak{S} (\rho_S (s), \sigma_S (s))$ with $\mathfrak{S} (\rho_S (s)
\otimes \rho_E (s), \sigma_S (s) \otimes \rho_E (s))$ by virtue of
Eq.~(\ref{eq:trinv}), thus getting
\begin{eqnarray}
  \Delta_S \mathfrak{S} (t, s) & \leqslant & \mathfrak{S} (\rho_{SE} (s),
  \sigma_{SE} (s)) \\
  &  & -\mathfrak{S} (\rho_S (s) \otimes \rho_E (s), \sigma_{SE}
  (s)) \nonumber\\
  &  & +\mathfrak{S} (\rho_S (s) \otimes \rho_E (s), \sigma_{SE} (s)) \nonumber\\
  &  & -\mathfrak{S} (\rho_S (s) \otimes \rho_E (s), \sigma_S (s) \otimes
  \rho_E (s)) . \nonumber
\end{eqnarray}
Applying the triangle-like inequalities, respectively, (\ref{eq:tlike2}) to
the first two terms at the r.h.s. of the previous expression and
(\ref{eq:tlike1}) to the last two terms, we get
\begin{eqnarray}
  \Delta_S \mathfrak{S} (t, s) & \leqslant & \phi (\mathfrak{S}(\rho_{SE} (s),
  \rho_S (s) \otimes \rho_E (s))) \nonumber\\
  &  & + \phi (\mathfrak{S}(\sigma_{SE} (s), \sigma_S (s) \otimes \rho_E
  (s))) .  \label{eq:almost}
\end{eqnarray}
Using once again Eq.~(\ref{eq:tlike1}), we have
\begin{eqnarray}
  &  & \mathfrak{S} (\sigma_{SE} (s), \sigma_S (s) \otimes \rho_E (s))
  \leqslant \mathfrak{S} (\sigma_{SE} (s), \sigma_S (s) \otimes \sigma_E (s))\label{eq:almost2}
  \\
  &  & \hphantom{pippo pippo}+ \phi (\mathfrak{S}(\sigma_S (s) \otimes \rho_E (s), \sigma_S (s)
  \otimes \sigma_E (s))) \nonumber \\
  &  & \hphantom{pippo}=\mathfrak{S} (\sigma_{SE} (s), \sigma_S (s) \otimes \sigma_E (s)) +
  \phi (\mathfrak{S}(\rho_E (s), \sigma_E (s))),   \nonumber
\end{eqnarray}
where in the equality we used Eq.~(\ref{eq:trinv}). The last step of the proof
follows from the fact that since $\phi$ is a concave non-negative function on
non-negative real numbers ($\mathfrak{S} \geqslant 0$ due to property I.) such
that $\phi (0) = 0$, then $\phi$ is also monotonically non-decreasing and
subadditive {\cite{DePonti2020a}}, so that Eq.~(\ref{eq:almost2}) implies
\begin{eqnarray}
  &  & \phi (\mathfrak{S}(\sigma_{SE} (s), \sigma_S (s) \otimes \rho_E (s))) 
  \label{eq:almost3}\\
  &  & \leqslant \phi (\mathfrak{S}(\sigma_{SE} (s), \sigma_S (s) \otimes
  \sigma_E (s))) + \phi \circ \phi (\mathfrak{S}(\rho_E (s), \sigma_E (s))),  \nonumber
\end{eqnarray}
which replaced in Eq.~(\ref{eq:almost}) directly leads us to the wanted
Eq.~(\ref{eq:main}).

Note that Eq.~(\ref{eq:main}) only depends on the defining properties I.-III.;
yet, it can be possible to derive alternative bounds to $\Delta \mathfrak{S}_S
(t, s)$ depending on specific choices of $\mathfrak{S}$, as will be
exemplified in the following. As we will see in the next sections, in the
considered cases the triangle-like inequalities build upon the validity of
inequalities of the form
\begin{eqnarray}
  D^2 (\rho, \sigma) & \leqslant & k \hspace{0.17em} \mathfrak{S} (\rho,
  \sigma)  \label{eq:pins}
\end{eqnarray}
with $k$ a positive coefficient and $D (\rho, \sigma)$ the trace distance
between $\rho$ and $\sigma$, defined as
\begin{equation}
  \label{eq:td} D (\rho, \sigma) = \frac{1}{2}  \| \rho - \sigma \|_1 =
  \frac{1}{2}  \sum_i | \ell_i |,
\end{equation}
where $\| \cdummy \|_1$ is the $1 -$norm, so that the $\ell_i$s are the
eigenvalues of the traceless operator $\rho - \sigma$.

In the remainder of the paper, we give significant examples of
distinguishability quantifiers representing specific instances of the general
framework defined here.

\section{Holevo skew divergence}\label{sec:holevo-skew-diverg}

Let us first introduce a quantum divergence directly derived from the
Holevo quantity, {thus establishing a clear link between non-Markovianity in terms of information backflow and a quantity of central interest in quantum information, communication and computation \cite{Nielsen2000}. The Holevo quantity associated with an ensemble of quantum states, each prepared with a certain probability,
tells us how much the von-Neumann entropy of the ensemble 
is reduced on average when we know which state of the ensemble 
has been prepared. If we now consider in particular an ensemble of two states, representing
two possible initial conditions of an open-system dynamics, and we follow the evolution
of the corresponding Holevo quantity, any increase in a given time interval 
means that the information gained, on average, by
knowing which initial state has been prepared would actually increase during that time interval. 
Thus, the Holevo quantity is a natural candidate to identify non-Markovianity
with the presence of time intervals of the dynamics of the open system 
during which the latter recovers some information that was previously flown to the environment.
As we are now going to show, this picture can be put on a firm ground within the theoretical
framework described in Sec.\ref{sec:cfn}. } 

{Given two states $\rho$ and
$\sigma$ and a mixing parameter $\mu$, with $0 < \mu < 1$, the
Holevo quantity restricted to a two-state ensemble $\{ \mu, \rho ; 1 - \mu,
\sigma \}$ takes the form
\begin{equation}
  \chi_{\mu} (\rho, \sigma) = S (\mu \rho + (1 - \mu) \sigma) - \mu S (\rho) -
  (1 - \mu) S (\sigma),
\end{equation}
with $S (\rho) = - \text{tr} \{ \rho \log \rho \}$ the von-Neumann entropy (note that we excluded the values $\mu=0,1$ which would  lead to the null quantity)}.
Now, since $0 \leqslant \chi_{\mu} (\rho, \sigma) \leqslant h (\mu)$, where
\begin{equation}
  h (\mu) = - \mu \log \mu - (1 - \mu) \log (1 - \mu)
\end{equation}
is the Shannon entropy of the probability distribution $\{ \mu, 1 - \mu \}$,
we define the quantity
\begin{equation}
  K_{\mu} (\rho, \sigma) = \frac{\chi_{\mu} (\rho, \sigma)}{h (\mu)}
\end{equation}
that is bounded between 0 and 1, and it is equal to 0 if and only if $\rho =
\sigma$, while it is equal to 1 if and only if $\rho$ and $\sigma$ have
orthogonal support. 

Hence, $K_{\mu} (\rho, \sigma)$ satisfies the property I.,
and we will see that it also satisfies properties II. and III., thus being a
quantum divergence according to our definition. We thus name $K_{\mu}$ Holevo
skew divergence, where the word skew refers to the fact that $\mu$ can be seen
as a skewing parameter that fixes the mixing of the two states $\rho$ and
$\sigma$ defining the divergence, while the term divergence stresses the fact
that the quantity only depends on two states and can therefore be taken as a
distinguishability quantifier, though it is not a distance. Finally, we note that the factor $(h (\mu))^{-
1}$ in the expression of the Holevo skew divergence, besides ensuring
normalization, makes $K_{\mu} (\rho, \sigma)$ independent from the logarithm
base used in its definition.

\subsection{Contractivity and Pinsker-like inequality}

The Holevo skew divergence inherits several important properties from its
connection with the quantum relative entropy. 
{The quantum relative entropy is generally defined for a pair of non-negative operators $A,B$ as 
\begin{equation}\label{eq:relativageneral}
  S (A ,B) =\text{tr} \{ A \log A \} - \text{tr} \{ A \log B \}+
  \text{tr} (B - A),
\end{equation}
that is a positive and finite quantity, provided that the support of $B$ includes the support of $A$ (where the convention $0\log(0)=0$ is  used), while it is defined to be infinity otherwise.
For a pair of statistical operators it therefore takes the more familiar form
{\cite{Bengtsson2017}}}
\begin{equation}\label{eq:relativa}
  S (\rho, \sigma) = \text{tr} \{ \rho \log \rho \} - \text{tr} \{ \rho \log
  \sigma \},
\end{equation}
so that we have in fact
\begin{eqnarray}
  K_{\mu} (\rho, \sigma) & = & \frac{\mu}{h (\mu)} S (\rho, \mu \rho + (1 -
  \mu) \sigma) \nonumber\\
  &  & + \frac{1 - \mu}{h (1 - \mu)} S (\sigma, (1 - \mu) \sigma + \mu \rho)
  .  \label{eq:holrel}
\end{eqnarray}
Indeed, the quantum relative entropy diverges whenever $\rho$ and $\sigma$
have orthogonal support; the Holevo skew divergence can thus be seen as a way
to regularize the quantum relative entropy to ensure boundedness and obtain an
entropic distinguishability quantifier. Importantly, in accordance with this
interpretation the Holevo skew divergence is symmetric under permutation of
the elements of the ensemble, as it immediately appears in \
Eq.~(\ref{eq:holrel}), so that
\begin{equation}
  K_{\mu} (\rho, \sigma) = K_{1 - \mu} (\sigma, \rho) . \label{eq:hsimm}
\end{equation}
In particular, the contractivity of the quantum relative entropy, $S (\Lambda
[\rho], \Lambda [\sigma]) \leqslant S (\rho, \sigma)$ for any CPTP map,
directly implies the contractivity of the Holevo skew divergence for any
parameter $\mu$https://it.overleaf.com/project/61c21cb03cbd4fb3b7efc7ff
\begin{equation}
  K_{\mu} (\Lambda [\rho], \Lambda [\sigma]) \leqslant K_{\mu} (\rho, \sigma),
\end{equation}
as can be readily seen by Eq.~(\ref{eq:holrel}) and the linearity of the map
$\Lambda$; in other terms, $K_{\mu}$ satisfies also the property II. expressed
by Eq.~(\ref{eq:cp}). Actually, the quantum relative entropy, and thus the
Holevo skew divergence as well, is contractive under maps that are simply
positive and trace preserving, but not necessarily CPTP {\cite{Hermes2013}}.

A further property that the Holevo skew divergence inherits from the quantum
relative entropy and that will be crucial for our purposes is the possibility
to lower bound it with the square of the trace distance, by means of an
inequality as in Eq.~(\ref{eq:pins}). {Starting from the Pinsker inequality for
the quantum relative entropy {\cite{Fumio1981a,Fedotov2003a,Bengtsson2017}}}
\begin{equation}
  D^2 (\rho, \sigma) \leqslant \frac{1}{2} S (\rho, \sigma)
\end{equation}
and using Eq.~(\ref{eq:holrel}), along with
\begin{eqnarray}
  D (\rho, \mu \rho + (1 - \mu) \sigma) & = & (1 - \mu) D (\rho, \sigma),
  \nonumber\\
  D (\sigma, (1 - \mu) \sigma + \mu \rho) & = & \mu D (\rho, \sigma), 
\end{eqnarray}
we find
\begin{equation}
  \label{eq:pinshol} D^2 (\rho, \sigma) \leqslant \frac{h
  (\mu)}{2 \mu (1 - \mu)} K_{\mu} (\rho, \sigma) .
\end{equation}
This relation represents an application of the Pinsker inequality to a
different entropic quantifier of state distinguishability and we will thus
refer to it as Pinsker-like inequality. Most importantly, it allows us to show
that the Holevo skew divergence satisfies also the property III. and
then to conclude that it is a proper 
quantifier of the information exchange between an open quantum system and its environment.

\subsection{Quantifier of information flow}

To prove the triangle-like inequalities in Eqs.(\ref{eq:tlike1}) and
(\ref{eq:tlike2}) for the Holevo skew divergence, we can exploit once again
its connection with the quantum relative entropy, along with the following
property of the latter. Given any three positive  operators $W, X, Y$, one has
{\cite{Audenaert2014a, audenaert2011telescopic}} 
\begin{eqnarray}\label{eq:essence1}
  &  & 0 \leqslant S (W, W + X) - S (W, W + X + Y) \nonumber\\
  &  & \hphantom{0} \leqslant \text{tr} W \log \left( 1+\frac{\text{tr}Y}{\text{tr} W} \right), \\
  \label{eq:essence2}
  &  & 0 \leqslant S (X, X + W) - S (X + Y, X + Y + W) \nonumber\\
  &  & \hphantom{0} \leqslant \text{tr} Y \log \left( 1+\frac{\text{tr}W}{\text{tr} Y} \right). 
\end{eqnarray}
As shown in Appendix \ref{app:pr}, these inequalities imply that given two
quantum relative entropies, each involving one of the two distinct states $\rho_1$
and $\rho_2$ together with the mixture with the same state $\sigma$ via the
same mixing parameter $\mu$, {which takes value in $(0,1)$,} their difference is bounded by
\begin{eqnarray}
  &  & S (\sigma, \mu \sigma + (1 - \mu) \rho_1) - S (\sigma, \mu \sigma + (1
  - \mu) \rho_2) \nonumber\\
  &  & \leqslant \log \left( 1 + \frac{1 - \mu}{\mu} D (\rho_1, \rho_2)
  \right) ;  \label{eq:b1}
\end{eqnarray}
analogously, it also holds
\begin{eqnarray}
  &  & S (\rho_1, \mu \rho_1 + (1 - \mu) \sigma) - S (\rho_2, \mu \rho_2 + (1
  - \mu) \sigma) \nonumber\\
  &  & \leqslant D (\rho_1, \rho_2) \log \left( 1 + \frac{1 - \mu}{\mu} 
  \frac{1}{D (\rho_1, \rho_2)} \right).  \label{eq:b2}
\end{eqnarray}
The terms at the left hand side (l.h.s.) of the previous inequalities are precisely of
the form of the terms connecting the Holevo skew divergence and the quantum
relative entropy in Eq.~(\ref{eq:holrel}), so that we immediately get
\begin{eqnarray}
  K_{\mu} (\rho, \sigma) - K_{\mu} (\rho, \tau) & \leqslant & g_{\mu} (D
  (\sigma, \tau)),  \label{eq:g}
\end{eqnarray}
where we introduced the function
\begin{eqnarray}
  g_{\mu} (x) & = & \frac{\mu}{h (\mu)} \log \left( 1 + \frac{1 - \mu}{\mu} x
  \right) \nonumber\\
  &  & + \frac{1 - \mu}{h (\mu)} x \log \left( 1 + \frac{\mu}{1 - \mu} 
  \frac{1}{x} \right) . 
\end{eqnarray}
Further using
\begin{equation}
  g_{\mu} (x) \leqslant \frac{\sqrt{4 \mu (1 - \mu)}}{h (\mu)}  \sqrt{x},
  \label{eq:sqrt}
\end{equation}
which  follows from the approximation $\log (1 + x) \leqslant \sqrt{x}$,
and the Pinsker-like inequality in Eq.~(\ref{eq:pinshol}), we finally get the
inequality
\begin{equation}
  \label{eq:tlike1hol} K_{\mu} (\rho, \sigma) - K_{\mu} (\rho, \tau) \leqslant
  \kappa_{\mu}  \sqrt[4]{K_{\mu} (\sigma, \tau)}
\end{equation}
with
\begin{equation}
  \label{eq:cu} \kappa_{\mu} = \sqrt[4]{\frac{8 \mu (1 - \mu)}{h^3
  (\mu)}} .
\end{equation}
This is indeed a triangle-like inequality as in Eq.~(\ref{eq:tlike1}), for the
concave function
\begin{equation}
  \phi_{\mu} (x) = \kappa_{\mu}  \sqrt[4]{x} \label{eq:phimuk}
\end{equation}
satisfying $\phi_{\mu} (x) > 0$ for $x > 0$ and $\phi (0) = 0$. In addition,
thanks to Eq.~(\ref{eq:hsimm}) and $\kappa_{\mu} = \kappa_{1 - \mu}$,
we have that Eq.~(\ref{eq:g}) implies also
\begin{equation}
  \label{eq:tlike2hol} K_{\mu} (\sigma, \rho) - K_{\mu} (\tau, \rho) \leqslant
  \kappa_{\mu}  \sqrt[4]{K_{\mu} (\sigma, \tau)},
\end{equation}
which is the triangle-like inequality in Eq.~(\ref{eq:tlike2}) with respect to
the given concave function $\phi_{\mu} (x)$.

We have thus shown that the Holevo skew divergence does satisfy all the
required properties I.-III. We can therefore apply to it the general picture
introduced in Sec.\ref{sec:ieb} to characterize the information flow in open
quantum system dynamics. Explicitly, the changes of information within the
open system are quantified by the variation of the Holevo skew divergence
according to Eq.~(\ref{eq:main}) with $\mathfrak{S} \rightarrow K_{\mu}$ and
$\phi$ given by Eq.~(\ref{eq:phimuk}). As shown in the Appendix \ref{app:ti},
this result can be improved exploiting directly the triangle inequality for
the trace distance in Eq.~(\ref{eq:g}) together with subadditivity of the
square root approximation of $g_{\mu}$ given by Eq.~(\ref{eq:phimuk}), thus
coming to
\begin{align}
  \Delta_S K_{\mu} (t, s) \leqslant \kappa_{\mu} &\left( \sqrt[4]{K_{\mu}
  (\rho_E (s), \sigma_E (s))} \right.  \label{eq:tighol}\\
  &\hphantom{it}+ \sqrt[4]{K_{\mu} (\rho_{SE} (s), \rho_S (s) \otimes \rho_E (s))}
  \nonumber\\
    & \hphantom{it}\left. + \sqrt[4]{K_{\mu} (\sigma_{SE} (s), \sigma_S (s) \otimes
  \sigma_E (s))} \right) . \nonumber
\end{align}
The
information contained at time $s$ within the environment and in the
system-environment correlations here quantified via the Holevo skew divergence
is thus responsible for any possible subsequent enhancement of the open-system
state distinguishability, in turn quantified via $\Delta_S K_{\mu} (t, s)$. Interestingly, in this expression all the contributions to the information
content within the environment and the system-environment correlations are
equally weighted by the same fourth root function and the same constant factor
$\kappa_{\mu}$, which takes its mimimum value for $\mu = 1 / 2$. 

\section{Distances and divergences}\label{sec:dad}

Besides accounting for the Holevo skew divergence, our approach
connects within a common framework several distinct witnesses of quantum
non-Markovianity, based on both distance- and divergence-based quantifiers of
state distinguishability.

\subsection{Helstrom norm and trace distance}

Given two quantum states $\rho$ and $\sigma$, the Helstrom norm $D_{\mu}
(\rho, \sigma)$ is the $\| \cdot \|_1 -$norm of the Hermitian operator given
by the difference of the two states, weighted by $\mu$ and $1 - \mu$
respectively, i.e.,
\begin{equation}
  \label{eq:hel} D_{\mu} (\rho, \sigma) = \| \mu \rho - (1 - \mu) \sigma \|_1;
\end{equation}
note that $D_{\mu} (\rho, \sigma)$ satisfies the symmetry property
\begin{equation}
  D_{\mu} (\rho, \sigma) = D_{1 - \mu} (\sigma, \rho) \label{eq:dsimm}
\end{equation}
as in Eq.~(\ref{eq:hsimm}). This quantity fixes the maximum success
probability in discriminating among $\rho$ and $\sigma$, if they have been
prepared with probability $\mu$ and $1 - \mu$ {\cite{Helstrom1969}}. Relying
on this, the Helstrom norm has been used to quantify the information flow in
open quantum system dynamics and to define accordingly a measure of quantum
non-Markovianity {\cite{Wissmann2015,Breuer2016,Amato2018a}}. 
{Quite
interestingly, the definition of quantum Markovian dynamics expressed via the
Helstrom norm under the assumption that the dynamical maps
are invertible turns out to be equivalent to the P-divisibility of the dynamics 
\cite{Vacchini2011,Chruscinski2011a,Wissmann2015,Chruscinski2018a}}, i.e., the possibility to decompose the dynamical maps
$\Lambda (t)$ as
\begin{equation}
  \Lambda (t) = \Lambda (t, s) \Lambda (s) \quad \forall \hspace{0.17em} t
  \geqslant s \geqslant 0,
\end{equation}
where $\Lambda (t, s)$ are positive (but not necessarily completely positive)
maps. The trace distance, see Eq.~(\ref{eq:td}), represents the specific
instance of the Helstrom norm for $\mu = 1 / 2$, $D (\rho, \sigma) = D_{1 / 2}
(\rho, \sigma)$, which is associated with the unbiased discrimination scenario
where the two states $\rho$ and $\sigma$ have been prepared with equal
probability. The trace-distance based definition of quantum Markovianity
{\cite{Breuer2009,Laine2010}} is the prototypical definition relying on the
notion of information flow and, more in general, the corresponding
non-Markovianity measure is one of the most significant quantifiers of quantum
non-Markovianity {\cite{Breuer2016}}. The approach via this generalized trace
distance, just thanks to this relation to P-divisibility, further allows to
make a connection to classical Markovian stochastic processes
{\cite{Wissmann2015}}. It moreover allows to overcome one of the criticism
against the trace distance approach, which is not sensitive to the action of
non-unital maps {\cite{Liu2013b}}.

The trace norm immediately satisfies the properties I.-III. defining a quantum
divergence and allowing us to apply the general framework introduced in
Sec.\ref{sec:cfn}, since it is a distance contractive under CPTP maps;
actually, also the trace distance is contractive under the weaker assumption
of positivity. Moving to the general Helstrom norm in Eq.~(\ref{eq:hel}), it is
convenient to consider its corresponding symmetrized version, that is,
\begin{equation}
  H_{\mu} (\rho, \sigma) = \frac{1}{2}  (D_{\mu} (\rho, \sigma) + D_{\mu}
  (\sigma, \rho)) .
\end{equation}
Clearly $H_{\mu} (\rho, \sigma)$ inherits the contractivity under (C)PTP maps
from the Helstrom norm {\cite{Wissmann2015,Breuer2016}}, while using the
triangle inequality and its reverse for the $\| \cdot \|_1 -$norm, $\|A\|_1 -
\|B\|_1 \leqslant \|A \pm B\|_1 \leqslant \|A\|_1 + \|B\|_1$, it is \ easy to
see that $H_{\mu} (\rho, \sigma)$ is lower bounded by the trace-distance, $D
(\rho, \sigma) \leqslant H_{\mu} (\rho, \sigma)$, while
\begin{eqnarray}
  D_{\mu} (\rho, \sigma) - D_{\mu} (\rho, \tau) & \leqslant & 2 (1 - \mu) D
  (\sigma, \tau), \nonumber\\
  D_{\mu} (\sigma, \rho) - D_{\mu} (\tau, \rho) & \leqslant & 2 \mu D (\sigma,
  \tau), 
\end{eqnarray}
which combined together lead to the triangle inequality
\begin{equation}
  \label{eq:trhel} H_{\mu} (\rho, \sigma) - H_{\mu} (\rho, \tau) \leqslant
  H_{\mu} (\sigma, \tau),
\end{equation}
and therefore to Eq.~(\ref{eq:main}) with $\phi (x) = x$. By direct inspection
as shown in {\cite{Amato2018a}} one also has the bound
\begin{eqnarray}
  \Delta_S D_{\mu}  (t, s) & \leqslant & 2 \min \{ \mu, 1 - \mu \} D (\rho_E
  (s), \sigma_E (s))  \label{eq:maintred}\\
  &  & + 2 \mu D (\rho_{SE} (s), \rho_S (s) \otimes \rho_E (s)) \nonumber\\
  &  & + 2 (1 - \mu) D (\sigma_{SE} (s), \sigma_S (s) \otimes \sigma_E (s)) .
  \nonumber
\end{eqnarray}

\subsection{Quantum skew divergence}

More recently {\cite{Megier2021}}, it has been shown that a full
characterization of quantum non-Markovianity in terms of a bidirectional
exchange of information between the open system and the environment can be
given in terms of entropic quantities, which, in particular, do not satisfy
the triangle inequality. We now show that also these quantities are
included into the general framework here introduced.

Let us define the quantum skew divergence as follows
\begin{eqnarray}
  S_{\mu} (\rho, \sigma) & = & \frac{\mu}{\log (1 / \mu)} S (\rho, \mu \rho +
  (1 - \mu) \sigma) \nonumber\\
  &  & + \frac{1 - \mu}{\log (1 / (1 - \mu))} S (\sigma, (1 - \mu) \sigma +
  \mu \rho),  \label{eq:tre}
\end{eqnarray}
with skewing parameter $\mu \in (0, 1)$.
{Note that each term is
finite for arbitrary $\mu$ and arbitrary pair of quantum states
$\rho$ and $\sigma$, pure or mixed, at variance with the quantum relative
entropy.}
This quantity is based on the telescopic relative entropy or quantum skew
divergence introduced in {\cite{Audenaert2014,Audenaert2014a, Megier2021}},
albeit with a symmetrization with respect to the simultaneous exchange $\mu
\leftrightarrow 1 - \mu$ and $\rho \leftrightarrow \sigma$, so that
\begin{equation}
  S_{\mu} (\rho, \sigma) = S_{1 - \mu} (\sigma, \rho), \label{eq:ssimm}
\end{equation}
which makes it a natural distinguishability quantifier. 
{In fact, $S_{\mu} (\rho, \sigma)$ provides us with a regularized and symmetrized version
of the relative entropy, that is a fundamental quantifier of the distinguishability
of quantum states and has been studied as a possible identifier of 
memory effects since the very beginning of the investigations on quantum non-Markovianity~\cite{Laine2010}.
The general framework presented in Sec.\ref{sec:cfn} allows us to provide also 
the regularized and symmetrized relative entropy with a complete interpretation in terms
of a quantifier of the information exchange between the open system and the environment.

The quantum skew divergence defined in Eq.(\ref{eq:tre})} satisfies the property I. of quantum divergences, i.e.,
$0 \leqslant S_{\mu}  (\rho, \sigma) \leqslant 1$, with the lower and upper
bounds being saturated if and only if $\rho = \sigma$ and {$\rho \perp_{\mbox{\footnotesize{supp}}}\sigma$},
respectively; 
indeed, $S_{\mu}  (\rho, \sigma)$ is independent from the
logarithm base in its definition by virtue of the normalizing prefactor
inversely proportional to the logarithm. In addition, the quantum skew divergence satisfies a Pinsker-like inequality, see Eq.~(\ref{eq:pins}) and
compare with Eq.~(\ref{eq:pinshol}), that reads
\begin{equation}
  \label{eq:pinstre} D^2 (\rho, \sigma) \leqslant \frac{\log
  (\mu) \log (1 - \mu)}{2 \mu (1 - \mu) h (\mu)} S_{\mu}  (\rho, \sigma) .
\end{equation}
Using this inequality, along with Eqs.(\ref{eq:b1}) and (\ref{eq:b2}) leading
to
\begin{eqnarray}
  S_{\mu} (\rho, \sigma) - S_{\mu} (\rho, \tau) & \leqslant & f_{\mu} (D
  (\sigma, \tau)),  \label{eq:f}
\end{eqnarray}
where we introduced the function
\begin{eqnarray}
  f_{\mu} (x) & = & \frac{\mu}{\log (1 / \mu)} \log \left( 1 + \frac{1 -
  \mu}{\mu} x \right) \nonumber\\
  &  & + \frac{1 - \mu}{\log (1 / (1 - \mu))} x \log \left( 1 + \frac{\mu}{1
  - \mu}  \frac{1}{x} \right), 
\end{eqnarray}
and using again the approximation $\log (1 + x) \leqslant \sqrt{x}$, we come
to
\begin{eqnarray}
  S_{\mu} (\rho, \sigma) - S_{\mu} (\rho, \tau) & \leqslant & \varsigma_{\mu} 
  \sqrt[4]{S_{\mu} (\sigma, \tau)} \\
  S_{\mu} (\sigma, \rho) - S_{\mu} (\tau, \rho) & \leqslant & \varsigma_{\mu} 
  \sqrt[4]{S_{\mu} (\sigma, \tau)} 
\end{eqnarray}
with
\begin{equation}
  \varsigma_{\mu} = \log \left( \frac{1}{\mu (1 - \mu)} \right)
  \sqrt[4]{\frac{\mu (1 - \mu)}{2\, h (\mu) \log^3  (\mu)
  \log^3  (1 - \mu)}} ;
\end{equation}
these are indeed triangle-like inequalities as in Eqs.(\ref{eq:tlike1}) and
(\ref{eq:tlike2}) for the concave function $\varsigma_{\mu}  \sqrt[4]{x}$. As
for the Holevo skew divergence, the inequalities are fixed by the concave
function given by the fourth root, which is however multiplied by a different
factor; compare with Eqs.(\ref{eq:tlike1hol}), (\ref{eq:cu}) and
(\ref{eq:tlike2hol}). Both $\kappa_{\mu}$ and $\varsigma_{\mu}$ due to the
symmetric choice reach their minimum value for $\mu = 1 / 2$,
corresponding to $\sqrt[4]{2/ \log^3  (2)} \approx 1.565$.

Finally, the quantum skew divergence inherits the contractivity under
(C)PTP maps from the quantum relative entropy, thus satisfying all the
defining properties of quantum divergences. Applying Eq.~(\ref{eq:main}), we
thus arrive at the upper bound with the usual interpretation in terms of
information flow from and toward the open system, linked to the information
within the environment and the system-environment correlations, now quantified
via the quantum skew divergence. Also in this case, a different and
tighter bound can be derived by using the Pinsker-like inequality
(\ref{eq:pinstre}) at a different stage of the derivation, in close analogy to
the calculations in Appendix \ref{app:ti}, thus obtaining
\begin{align}
  \Delta_S S_{\mu} (t, s) \leqslant \varsigma_{\mu} &\left( \sqrt[4]{S_{\mu}
  (\rho_E (s), \sigma_E (s))} \right.  \label{eq:maintre}\\
  &\hphantom{it}+ \sqrt[4]{S_{\mu} (\rho_{SE} (s), \rho_S (s) \otimes \rho_E (s))}
  \nonumber\\
    & \hphantom{it}\left. + \sqrt[4]{S_{\mu} (\sigma_{SE} (s), \sigma_S (s) \otimes
  \sigma_E (s))} \right) , \nonumber
\end{align}
which confirms the result obtained in {\cite{Megier2021}}, albeit with a
different symmetrization of the entropic distinguishability quantifier as
given by Eq.~(\ref{eq:tre}).

\subsection{Jensen-Shannon divergence}

As it immediately appears from the previous results, another significant
quantifier of state distinguishability and information flow is the
Jensen-Shannon divergence, which is defined starting from the quantum relative
entropy as
\begin{equation}
 \mathsf{J} (\rho, \sigma) = \frac{1}{2 \log 2}  \left( S \left( \rho, \frac{\rho +
  \sigma}{2} \right) + S \left( \sigma, \frac{\rho + \sigma}{2} \right)
  \right),
\end{equation}
where at variance with the usual definition {\cite{Bengtsson2017}} we have
considered a normalization factor such that the expression is independent from
the chosen logarithm basis and moreover it lies within the range $0 \leqslant
\mathsf{J} (\rho, \sigma) \leqslant 1$. Quite interestingly, with the considered
normalization the Jensen-Shannon divergence can be seen as the special
instance of the Holevo skew divergence as defined in Eq.~(\ref{eq:holrel}) \
for $\mu = 1 / 2$,
\begin{equation}
  \mathsf{J} (\rho, \sigma) = K_{1 / 2} (\rho, \sigma) .
\end{equation}
Equivalently, we can also recover the Jensen-Shannon divergence from the
quantum skew divergence as defined in Eq.~(\ref{eq:tre}), again setting $\mu =
1 / 2$
\begin{equation}
  \mathsf{J} (\rho, \sigma) = S_{1 / 2}  (\rho, \sigma) .
\end{equation}
As a consequence, we can directly see that the Jensen-Shannon divergence is a
quantum divergence according to our definition, and a bound on $\Delta_S \mathsf{J} (t,
s)$ can be readily derived from Eq.~(\ref{eq:main}) (or simply setting $\mu = 1
/ 2$ in either Eq.~(\ref{eq:tighol}) or Eq.~(\ref{eq:maintre})).

On the other hand, the square root of the Jensen-Shannon divergence $\sqrt{\mathsf{J}
(\rho, \sigma)}$ has been recently proven to be a distance
{\cite{Virosztek2021,Sra2021}}, satisfying in particular the triangle
inequalities, i..e., Eqs.(\ref{eq:tlike1}) and (\ref{eq:tlike2}) for $\phi (x)
= x$. The square root of the Jensen-Shannon divergence indeed still satisfies
boundedness, normalization and indistinguishability of identical states, as
well as the contractivity under (C)PTP map, due to the monotonicity of the
square root, thus providing us with a further example of quantum divergence.
Besides inequality Eq.~(\ref{eq:main}), the revival of the square
root of the Jensen-Shannon divergence can be bounded by the tighter inequality
{\cite{Megier2021}}
\begin{align}\label{eq:mainjs}
  \Delta_S  \sqrt{\mathsf{J} (t, s)} \leqslant & \sqrt{\mathsf{J} (\rho_E (s),
  \sigma_E (s))}  \\
  &  \hphantom{it}+ \sqrt{\mathsf{J} (\rho_{SE} (s), \rho_S (s) \otimes \rho_E (s))}  \nonumber \\
  &  \hphantom{it}  +
  \sqrt{\mathsf{J} (\sigma_{SE} (s), \sigma_S (s) \otimes \sigma_E (s))}  .
  \nonumber
\end{align}
As shown in the following example, the evolution of the square root of the
Jensen-Shannon divergence typically follows the evolution of the trace
distance, with respect to both the revivals of the open-system
distinguishability and the information content outside the open system, more
closely than the other quantifiers that are quantum divergences but not
distances.

\begin{figure*}
  \includegraphics[width=0.45\textwidth]{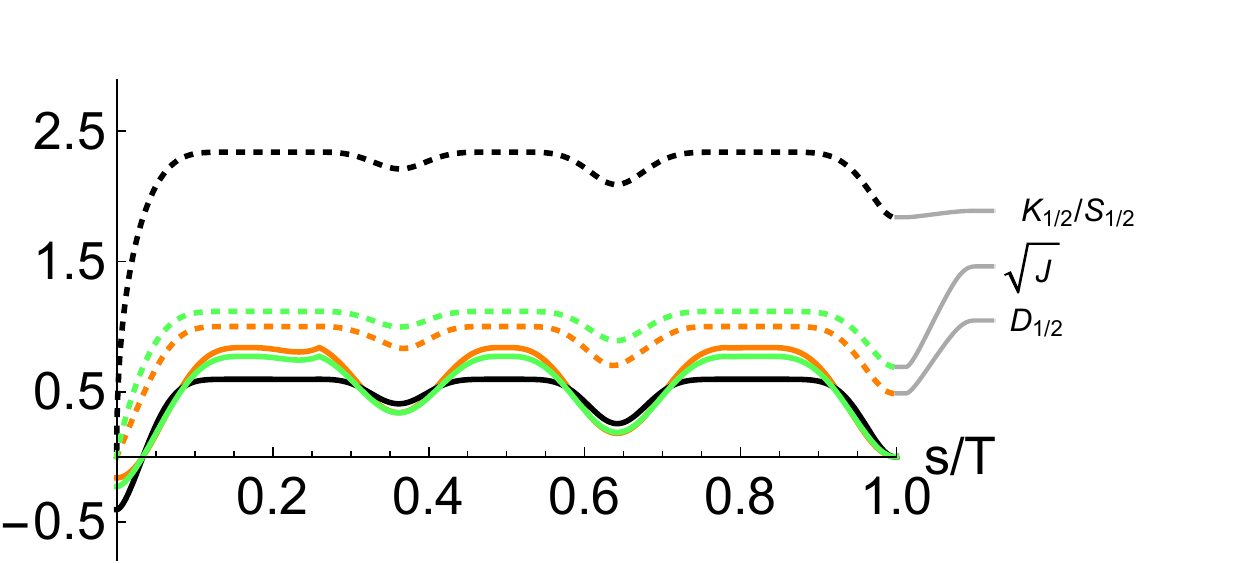}~~\includegraphics[width=0.45\textwidth]{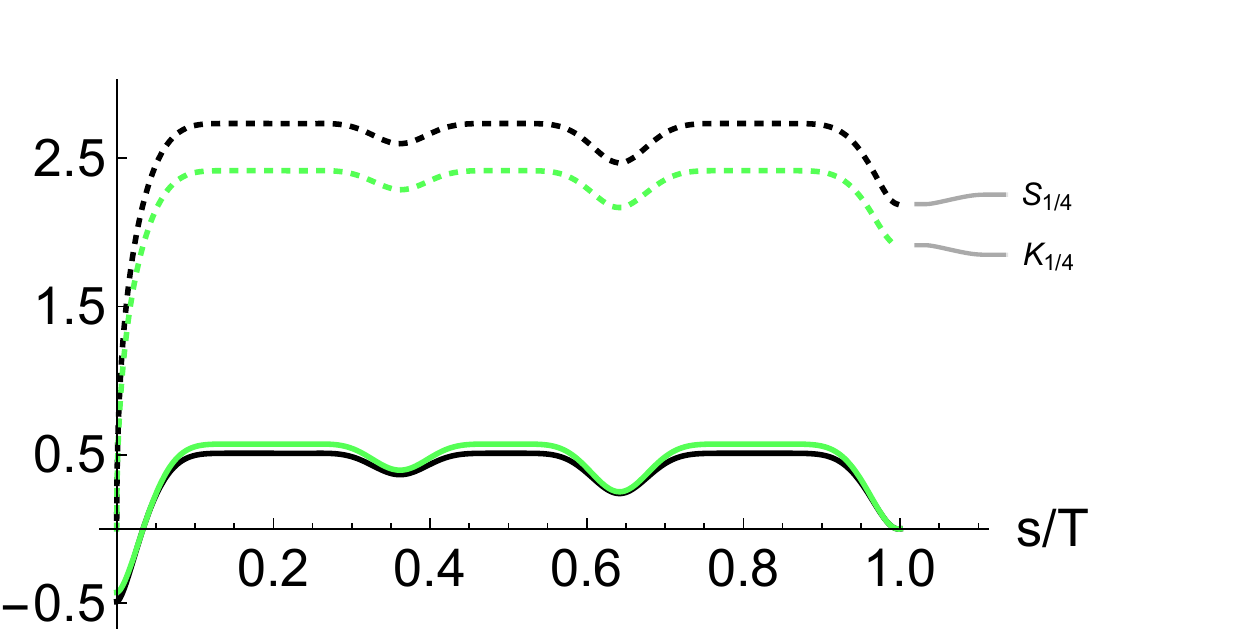}
  \includegraphics[width=0.45\textwidth]{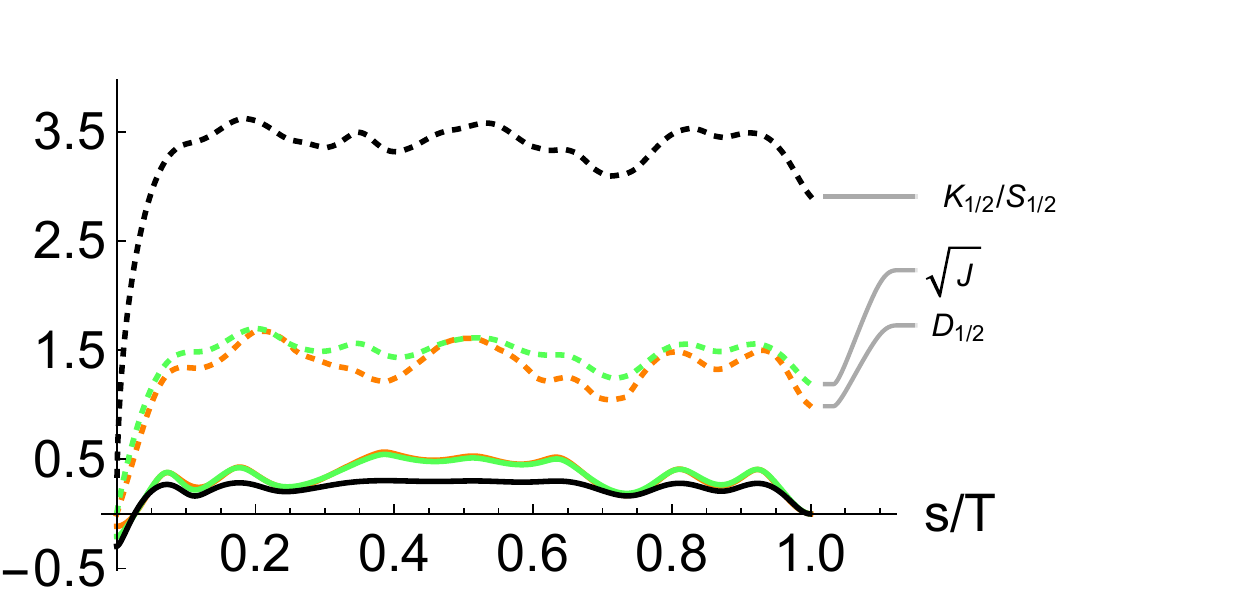}~~\includegraphics[width=0.45\textwidth]{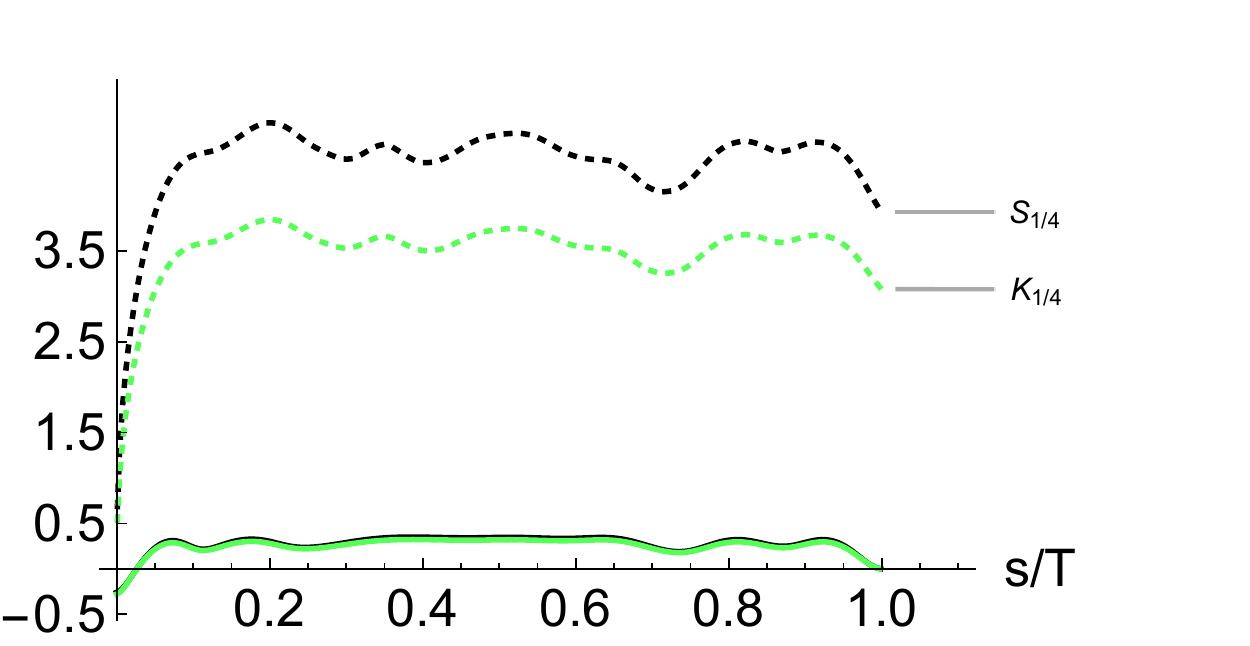}
\caption{%
  Revivals of distance and entropic distinguishability quantifiers
    versus their bounds for the two models considered in Sect.\ref{sec:ex}, namely the spin-star dephasing model with $N=5$
    environmental qubits and random coupling strengths (top panels) and the Jaynes-Cummings model (bottom panels). Left: Behavior
    of Helstrom norm [$D$, orange (light gray) lines], square root of the Jensen-Shannon
    divergence [$\sqrt{J}$, green (gray) lines], together with Holevo ($K$) and quantum ($S$)
    skew divergence evaluated for $\mu=1/2$ [black (dark gray) lines], so that the latter
    two coincide. The continuous lines represent the
    l.h.s. of Eq.~(\ref{eq:maintred}), Eq.~(\ref{eq:mainjs}) as well as
    Eq.~(\ref{eq:tighol}) and Eq.~(\ref{eq:maintre}) evaluated for $\mu=1/2$ and $t=T$, 
    respectively. The dotted lines denote the sum
    of the corresponding contributions at the r.h.s.. Right: Behavior
    of Holevo skew divergence [$K$, green (gray) lines] and quantum [$S$, black (dark gray) lines] 
    skew divergence
    for $\mu=1/4$. Again continuous lines correpond to the 
    l.h.s. of Eq.~(\ref{eq:tighol}) and Eq.~(\ref{eq:maintre}),
    respectively, while the dotted lines reproduce the r.h.s..  Despite the quite different nature of the two models, describing for a two-level system respectively decoherence, and excitation exchanges with a bosonic mode, the overall behavior of distinguishability quantifiers and related bounds is strikingly similar. In particular, it clearly appears that distance quantifiers provide tighter bounds. For the spin-star model the reference time $T$ is equal to 5 in inverse units of the
average coupling strength, while for the Jaynes-Cummings model $T$ is 8.9 in inverse units of the
coupling strength $g$, while the detuning frequency $\Delta$ is
equal to 0.5 in units of $g$.}
 \label{fig:ssjc}
\end{figure*}
\subsection{Role of skewing}\label{sec:role-skewing}

In Secs.\ref{sec:holevo-skew-diverg} and \ref{sec:dad} we have introduced different distinguishability
quantifiers which further qualify as non-Markovianity quantifiers according to
the properties I.-III. These properties allow for an interpretation of memory
effects as related to storage and retrival of information, in quantum degrees
of freedom not accessible by performing measurements on the system alone. All
these quantifiers are charaterized by a skewing parameter $\mu$. For the case
of entropic quantifiers, the introduction of this skewing parameter is necessary to introduce well-defined quantities and
avoid the divergences which plague the standard definition of quantum
relative entropy also in a finite dimensional setting. As shown by the present
analysis, different quantifiers can be introduced that all point to a
distinguished role of the value $1 / 2$ for the mixing parameter. This choice allows in particular to
obtain a distance from such entropic quantifiers, which however is
sensitive also to non-unital evolutions {\cite{Megier2021}}. It
remains open the question whether for these quantifiers a connection with
P-divisibility can also be established. For the case of the generalized trace
distance or Helstrom norm the value $\mu = 1/2$ for the skewing parameter
still plays a distinguished role, leading to recover the trace distance, but
it fails in dealing with non-unital dynamics. Moreover it does not allow any
more to identify contractivity and P-divisibility for invertible time
evolutions.

\section{Examples}\label{sec:ex}

\subsection{Spin-star configuration}
In order to exemplify the behavior of the different distinguishability
quantifiers, suitable for the description of non-Markovianity according
to properties  I.-III. of Sec.\ref{sec:cfn}, we first consider a model of $N$
qubits coupled to a reference qubit in the
so-called spin star configuration \cite{PhysRevA.76.052117,Breuer2004a},
according to the
Hamiltonian
\begin{equation}
H=\omega_{\scriptscriptstyle S} \sigma_z +\sum\limits_{k=1}^N g_k  \sigma_z\otimes  \sigma^k_z+ \sum\limits_{k=1}^N \omega^k_{\scriptscriptstyle E} \sigma^k_z,
\end{equation}
which describes a pure dephasing interaction;
such a model characterizes for example the reduced evolution of an electronic spin qubit 
in a diamond
nitrogen-vacancy center \cite{Torrontegui_2016,Haase2018}.
Here $\omega_{\scriptscriptstyle S}$ and $\omega^k_{\scriptscriptstyle E}$ denote the frequencies of the system and of the environmental qubits respectively, $\sigma_z$ is the Pauli matrix and the superscript $k$ labels the environmental units coupled with different strengths $g_k$.

For the considered choice of environment which starts in the maximally mixed
state, corresponding to a high temperature reservoir, the environment
is left unchanged also for a non-Markovian dynamics and the only
relevant contribution to the exchange of information between system
and environment is to be traced back to the establishment of
correlations.
For this model a natural choice of initial pair of states is given by
the orthogonal pure superposition states $(|1\rangle + |0\rangle)/\sqrt{2}$ and  $(|1\rangle - |0\rangle)/\sqrt{2}$, where $\{|1\rangle, |0\rangle\}$ denote the eigenstates of the $\sigma_z$ operator, corresponding to the reduced density matrices $\rho_+$ and $\rho_-$ respectively. For later times the reduced states take the form
\begin{align}
    \rho_\pm(t)=\frac{1}{2}\begin{pmatrix}
1 & \pm \prod_{k=1}^N  \cos(2g_kt) \\
\pm\prod_{k=1}^N \cos(2g_kt) & 1 
\end{pmatrix},
\end{align}
where the index $k$ is running over all environmental units. Note, that such a special choice of the initial reduced and environmental states does not influence the qualitative features of the left and right hand sides of inequalities (\ref{eq:tighol}), (\ref{eq:maintred}), (\ref{eq:maintre}) and (\ref{eq:mainjs}), depicted in Fig.~\ref{fig:ssjc}, associated with the Holevo skew divergence, symmetrized Helstrom norm, quantum skew divergence and square root of Jensen-Shannon divergence, respectively. On the top left panel, we have plotted the quantities for a skewing parameter $\mu=1/2$, in which case the Holevo and the quantum skew divergence coincide. We can strengthen the findings from \cite{Megier2021} and observe that though all of the quantities provide the same qualitative  picture, the two distances differ also quantitatively very little from each others. Remarkably, the two solid lines corresponding to the l.h.s. of the associated inequality almost overlap. On the other hand, the upper bounds given by proper quantum divergences (i.e. not distances) are much looser. For completeness, we have plotted on the top right panel of Fig.~\ref{fig:ssjc} the Holevo and the quantum skew divergence for a skewing parameter $\mu=1/4$, where the quantities, albeit now different, tightly follow each others; this is especially visible for the variations of the reduced quantities (solid lines).     
{Besides illustrating the different tightness of the bounds for the distinct
distinguishability quantifiers, Fig.~\ref{fig:ssjc} shows that the bounds follow qualitatively
the evolution of the corresponding quantifiers, reproducing in particular the subsequently enhanced and suppressed revivals of the information that is accessed by the open system in the course of time.}

\subsection{Jaynes-Cummings model}
As second case study, we consider a further model of physical interest, 
with ubiquitous applications for example in quantum optical systems, i.e., the Jaynes-Cummings model \cite{Walls1995}. Here, the open system is a two-level
system with transition frequency $\omega_{\scriptscriptstyle S}$, while the environment consists of a single bosonic mode of frequency $\omega_{\scriptscriptstyle E}$, with corresponding annihilation and
creation operators denoted as $b$ and $b^\dag$. The global Hamiltonian is
\begin{equation}\label{eq:hjc}
    H = \omega_{\scriptscriptstyle S} \sigma_+\sigma_- \otimes \mathbbm{1}_{\scriptscriptstyle E} + \omega_{\scriptscriptstyle E} \mathbbm{1}_{\scriptscriptstyle S} \otimes b^\dag b
    + g  \left(\sigma_+\otimes  b + \sigma_-\otimes  b^\dag \right),
\end{equation}
with $\sigma_{+} = \ket{1}\bra{0}$ and $\sigma_{-} = \ket{0}\bra{1}$ raising and lowering operators of the two-level system,
so that the interaction between the two-level system
and the mode preserves the total number of excitations.
The global unitary operator can be obtained exactly \cite{Puri2001} and thus the reduced dynamics
can be derived explicitly for fully general initial conditions \cite{Smirne2010,Smirne2010b}.
Introducing the functions of the number operator $\hat{n} = b^\dagger b$,
\begin{eqnarray}
  c \left( \hat{n}, t \right) & = & e^{i \Delta t / 2}  \left[ \cos \left(
  f(\hat{n})\frac{t}{2} \right)  - i \Delta \frac{\sin \left(f(\hat{n})
  \frac{t}{2} \right) }{f(\hat{n})}  \right], \nonumber\\
  d \left( \hat{n}, t \right) & = & - 2 i e^{i \Delta t / 2} g \frac{\sin
  \left(f(\hat{n})\frac{t}{2} \right) }{f(\hat{n})},  \label{eq:dn}
\end{eqnarray}
with $\Delta = \omega_{\scriptscriptstyle S}-\omega_{\scriptscriptstyle E}$ and
\begin{equation}
    f(\hat{n}) = \sqrt{\Delta^2 + 4 g^2 \hat{n}},
\end{equation}
the global unitary operator can be written in fact as
\begin{eqnarray}
  U (t) & = & 
    \ket{1}\bra{1}\otimes c_{} \left( \hat{n} + 1, t \right) + \ket{1}\bra{0}\otimes d \left( \hat{n} + 1, t \right) b \nonumber\\
&&    -  \ket{0}\bra{1}\otimes b^{\dag} d^{\dag} \left( \hat{n} + 1, t \right) + \ket{0}\bra{0}\otimes c^{\dag} \left(
    \hat{n}, t \right).  \label{eq:u}
\end{eqnarray}
In particular, for any initial product state $\rho_{\scriptscriptstyle S \scriptscriptstyle E}(0)= \rho_{\scriptscriptstyle S}(0) \otimes \rho_{ \scriptscriptstyle E}(0)$ with stationary initial environmental state (i.e., $[\rho_{\scriptscriptstyle E}(0), \hat{n}]=0$), the 
open-system state at time $t$ reads
\begin{equation}
    \rho_{\scriptscriptstyle S}(t) = \left(\begin{array}{cc}
   \rho_{00}\left( 1 -
  \alpha (t) \right) + \rho_{11}  \beta \left( t
  \right) &  \rho_{10} \gamma \left( t
  \right)\\
    \rho_{01} \gamma^{\ast}
  (t) & \rho_{00}  \alpha \left( t
  \right) + \rho_{11} \left( 1 - \beta (t)
  \right)
  \end{array}\right)\label{eq:rhosjct}
\end{equation}
where $\rho_{ij}$, $i,j=0,1$, denote indeed the initial reduced-state elements $\rho_{ij}=\bra{i}\rho_{\scriptscriptstyle S}(0)\ket{j}$,
and we introduced the time-dependent functions
\begin{eqnarray}
  \alpha (t) & = & \langle c^{\dag} \left( \hat{n}, t \right) c
  \left( \hat{n}, t \right) \rangle_E, \nonumber\\
  \beta (t) & = & \langle c^{\dag} \left( \hat{n} + 1, t \right)
  c \left( \hat{n} + 1, t \right) \rangle_E,  \label{eq:abc}\\
  \gamma (t) & = & \langle \mathbbm{} c \left( \hat{n}, t \right)
  c_{} \left( \hat{n} + 1, t \right) \rangle_E, \nonumber
\end{eqnarray}
with $\langle A \rangle_E = \mbox{tr}\left\{A \rho_{\scriptscriptstyle E}(0)\right\}$.
Then, Eq.(\ref{eq:rhosjct}) fully characterizes the open two-level system evolution and, in particular,
it determines the degree of non-Markovianity of the reduced dynamics; Eq.(\ref{eq:u}), on the other hand,
allows us to evaluate explicitly quantities referred to the global system, thus getting a complete description of 
the information exchange between the open system and the environment, via the quantifiers introduced in Sec.\ref{sec:dad}. 
The behavior of distance and entropic distinguishability quantifiers for the Jaynes-Cummings model is considered in the bottom panels of Fig.\ref{fig:ssjc}, 
considering as initial
states of the qubit the excited state and a
balanced superposition of excited and ground state,
while the environment starts in a thermal state with $\beta\omega_{\scriptscriptstyle E}=1$,
and essentially the same considerations made for the spin-star model apply. We stress that again the distance quantifiers almost overlap and exhibit tighter bounds with respect to the entropic quantifiers.

\section{Conclusions}\label{sec:con}
In this paper, {we have have provided a general framework to relate distinguishability quantifiers 
with the information exchange} between an open system and its environment.
In particular, besides normalization, indistinguishability of identical states
and contractivity under the action of CPTP maps, one needs triangle-like inequalities. Importantly, since the latter are weaker than the standard triangle inequality,
we could include in our analysis not only distances, but also quantum divergences that are not necessarily distances.
The mentioned properties directly lead to an upper bound of the distinguishability variations, which traces non-Markovianity back
to a flow of information from the system-environment correlations and the environment to the open system.

The general framework includes the Holevo skew divergence, that is a normalized version of the Holevo quantity, as a special instance. For this quantity we also derived a tighter upper bound, while keeping the same physical interpretation.
Moreover, we have compared our approach with the quantification of distinguishability via the Helstrom norm of the weighted difference of two quantum states, and we have shown that a regularized and symmetrized version of the relative entropy, i.e., the quantum skew divergence,
satisfies the defining properties of our general framework as well. Both the Holevo skew divergence and the quantum skew divergence 
reduce for the case of equal weights to the Jensen-Shannon divergence, whose square root
is a distance contractive under CPTP maps, thus also being part of the formalism defined here. 
All of these quantifiers are sensitive to non-unital dynamics for any value of the skewing parameter.
On the other hand, the Helstrom norm for the case of equal weights recovers the trace distance, which is left unaltered by all non-unital dynamics.

It remains to be clarified whether this approach can provide further insight on the relationship between the notion of non-Markovianity as due to information exchange, considered in this paper, and P-divisibility of the time evolution map.
{In addition, it will be worth investigating whether the class of system-environment information
quantifiers can be further extended, possibly leading to other upper bounds to the information revivals.}
Finally, we expect that our work can shed some light also on the investigations on the
relevance of different distinguishability quantifiers used in connection with
the detection of initial correlations as considered in
{\cite{Dajka2011a,Wissmann2013a}}.

\acknowledgments
All authors acknowledge
support from UniMi, via Transition Grant
H2020 and PSR-2 2020. NM was funded by the Alexander von Humboldt Foundation in form of a Feodor-Lynen Fellowship and project ApresSF, supported by the National Science Centre under the QuantERA programme, funded by the European Union's Horizon 2020 research and innovation programme.
The authors would like to thank the referees for the many valuable suggestions, including the study of the Jaynes-Cummings model.

\appendix

\section{Proof of the bounds in Eqs.(\ref{eq:b1}) and
(\ref{eq:b2})}\label{app:pr}
To prove Eq.~(\ref{eq:b1}) let us express the difference of quantum relative
entropies of interest exploiting their definition as in Eq.~(\ref{eq:relativa})
\begin{eqnarray}
  &  & S (\sigma, \mu \sigma + (1 - \mu) \rho_1) - S (\sigma, \mu \sigma + (1
  - \mu) \rho_2) \\
  &  & = \tmop{tr} \{ \sigma [\log (\mu \sigma + (1 - \mu) \rho_2) - \log
  (\mu \sigma + (1 - \mu) \rho_1)] \} \nonumber\\
  &  & = \tmop{tr} \left\{ \sigma \left[ \log \left( \sigma + \frac{1 -
  \mu}{\mu} \rho_2 \right) - \log \left( \sigma + \frac{1 - \mu}{\mu} \rho_1
  \right) \right] \right\} . \nonumber
\end{eqnarray}
Denoting with $T_+ (T_-)$ the positive (negative) part of a self-adjoint
operator $T$ so that
\begin{equation}
  T =  T_+ - T_-, 
\end{equation}
we can consider the simple inequality
\begin{eqnarray}\label{eq:app1}
  \rho_2 & = & \rho_1 + (\rho_2 - \rho_1) \\
  & = & \rho_1 + (\rho_2 - \rho_1)_+ - (\rho_2 - \rho_1)_- \nonumber\\
  & \leqslant & \rho_1 + (\rho_2 - \rho_1)_+ . \nonumber
\end{eqnarray}
Exploiting Eq.(\ref{eq:app1}) together with the operator monotonicity of the
logarithm and the inequality Eq.~(\ref{eq:essence1}) we obtain
\begin{eqnarray}
  &  & S (\sigma, \mu \sigma + (1 - \mu) \rho_1) - S (\sigma, \mu \sigma + (1
  - \mu) \rho_2) \\
  &  & \leqslant \tmop{tr} \left\{ \sigma \left[ \log \left( \sigma + \frac{1
  - \mu}{\mu}  \rho_1 + \frac{1 - \mu}{\mu} (\rho_2 - \rho_1)_+ \right) \right.\right.
       \nonumber
  \\
  & &
      \hphantom{nnnnn} - \left. \left.\log \left( \sigma + \frac{1 - \mu}{\mu} \rho_1 \right) \right]
  \right\} \nonumber\\
  &  & \leqslant \log \left( 1 + \frac{1 - \mu}{\mu} \tmop{tr} (\rho_2 -
  \rho_1)_+ \right), \nonumber
\end{eqnarray}
so that finally exploiting
\begin{equation}\label{eq:app2}
  D (\rho_1, \rho_2) = \frac{1}{2} \| \rho_1 - \rho_2 \| = \tmop{tr} (\rho_2 -
  \rho_1)_+= \tmop{tr} (\rho_2 -
  \rho_1)_-
\end{equation}
we have the desired bound Eq.~(\ref{eq:b1}).

In a similar way using Eq.~(\ref{eq:relativa}) we have
\begin{eqnarray}
  &  & S (\rho_1, \mu \rho_1 + (1 - \mu) \sigma) - S (\rho_2, \mu \rho_2 + (1
  - \mu) \sigma) \\
  &  & = S \left(\rho_1, \rho_1 + \frac{1 - \mu}{\mu} \sigma\right) - S \left(\rho_2, \rho_2
  + \frac{1 - \mu}{\mu} \sigma\right) \nonumber
\end{eqnarray}
so that making use of the fact that $(\rho_2 - \rho_1)_-$ is a positive operator
together with
\begin{eqnarray}
  S (\rho + w, \sigma + w) & \leqslant & S (\rho, \sigma) 
\end{eqnarray}
for positive $w$ we come to
\begin{eqnarray}
  &  & S (\rho_1, \mu \rho_1 + (1 - \mu) \sigma) - S (\rho_2, \mu \rho_2 + (1
  - \mu) \sigma) \\
  &  & \leqslant S \left(\rho_1, \rho_1 + \frac{1 - \mu}{\mu} \sigma\right) \nonumber \\
  &  & - S \left(\rho_2
  + (\rho_2 - \rho_1)_-, \rho_2 + (\rho_2 - \rho_1)_- + \frac{1 - \mu}{\mu}
  \sigma\right) \nonumber
\end{eqnarray}
and finally using Eq.~(\ref{eq:essence2}) so that
\begin{eqnarray}
  &  & S \left(\rho_2 + (\rho_2 - \rho_1)_-, \rho_2 + (\rho_2 - \rho_1)_- +
  \frac{1 - \mu}{\mu} \sigma\right) \\
  &  & \geqslant S \left(\rho_1, \rho_1 + \frac{1 - \mu}{\mu} \sigma\right) - \tmop{tr}
  \{ (\rho_2 - \rho_1)_- \} \log \left( 1 + \frac{\tmop{tr} \left\{ \frac{1 -
  \mu}{\mu} \sigma \right\}}{\tmop{tr} \{ (\rho_2 - \rho_1)_- \}} \right)
  \nonumber
\end{eqnarray}
we obtain further exploiting Eq.(\ref{eq:app2}) 
the final result Eq.~(\ref{eq:b2})
\begin{eqnarray}
  &  & S (\rho_1, \mu \rho_1 + (1 - \mu) \sigma) - S (\rho_2, \mu \rho_2 + (1
  - \mu) \sigma) \\
  &  & \leqslant D (\rho_1, \rho_2) \log \left( 1 + \frac{1 - \mu}{\mu}
  \frac{1}{D (\rho_1, \rho_2)} \right) . \nonumber
\end{eqnarray}

\section{Proof of the bound in Eq.(\ref{eq:tighol}) on the information flow via the Holevo skew
divergence}\label{app:ti}

As discussed in the main text, a direct application of the general framework
for the establishment of the connection between non-Markovianity and
information exchange between system and environment exposed in
Sec.\ref{sec:cfn} would lead directly to the bound
\begin{eqnarray}
  \Delta_S K_{\mu} (t, s) & \leqslant & \kappa^2_{\mu}  \sqrt[16]{K_{\mu}
  (\rho_E (s), \sigma_E (s))}  \\
  &  & + \kappa_{\mu}  \sqrt[4]{K_{\mu} (\rho_{SE} (s), \rho_S (s) \otimes
  \rho_E (s))} \nonumber\\
  &  & + \kappa_{\mu}  \sqrt[4]{K_{\mu} (\sigma_{SE} (s), \sigma_S (s)
  \otimes \sigma_E (s))} . \nonumber
\end{eqnarray}
It is actually possible to derive a different tighter upper bound to the
variation of the Holevo skew divergence, namely Eq.(\ref{eq:tighol}), for which the same physical
interpretation as the one above indeed applies. To this aim let us start from
Eq.~(\ref{eq:g}) and combine it with the upper bound Eq.~(\ref{eq:sqrt}) so that
we have
\begin{eqnarray}
  K_{\mu} (\rho, \sigma) - K_{\mu} (\rho, \tau) & \leqslant & \frac{\sqrt{4
  \mu (1 - \mu)}}{h (\mu)} \sqrt{D (\sigma, \tau)} .  
\end{eqnarray}
Starting from this inequality adding and subtracting terms we come as in
Eq.~(\ref{eq:almost}) to
\begin{eqnarray}
  \Delta_S K_{\mu} (t, s) \leqslant & \frac{\sqrt{4 \mu (1 - \mu)}}{h (\mu)}
  \left( \sqrt{D (\rho_{SE} (s), \rho_S (s) \otimes \rho_E (s))} 
\right.+\nonumber
  \\
  & + \left. \sqrt{D (\sigma_{SE} (s), \sigma_S (s) \otimes \rho_E (s))}
  \right) . 
\end{eqnarray}
We can now exploit the fact that the trace distance obeys the triangle
inequality, so that
\begin{eqnarray}
  D (\sigma_{SE} (s), \sigma_S (s) \otimes \rho_E (s))  \leqslant & D
  (\sigma_{SE} (s), \sigma_S (s) \otimes \sigma_E (s))  \nonumber\\
  &   + D (\sigma_E (s), \rho_E (s)),
\end{eqnarray}
together with subadditivity of the square root, thus coming to
\begin{multline}
  \Delta_S K_{\mu} (t, s) \leqslant  \frac{\sqrt{4 \mu (1 - \mu)}}{h (\mu)}
  \left( \sqrt{D (\rho_{SE} (s), \rho_S (s) \otimes \rho_E (s))} + \right. 
  \\ +\left. \sqrt{D (\sigma_{SE} (s), \sigma_S (s) \otimes \sigma_E (s))} +
  \sqrt{D (\sigma_E (s), \rho_E (s))} \right)  .
\end{multline}
At this stage we can apply the Pinsker-like inequality Eq.~(\ref{eq:pinshol})
to finally obtain Eq.~(\ref{eq:tighol}).

\end{document}